\documentclass{aa}

\begin{document}
\title{Enhanced mass transfer during dwarf nova outbursts by irradiation of
the secondary?}

\author{Yoji Osaki\thanks{Professor Emeritus, University of Tokyo and Nagasaki University}  \inst{1}
\and Friedrich Meyer \inst{2}}

\institute{Department of Astronomy, School of Science, University of Tokyo,
Bunkyo-ku, Tokyo, 113-0033, Japan
\and
  Max-Planck-Institut f\"ur Astrophysik, Karl-Schwarzschild-Str. 1,
D-85740 Garching, Germany}

\offprints{Yoji Osaki; osaki@astron.s.u-tokyo.ac.jp}

\date{Received: / Accepted:}

\abstract{
One of the remaining issues in the problems of dwarf novae is
whether
or not enhanced mass transfer due to irradiation of the secondary stars
could
occur during outbursts. In a previous paper (Osaki and Meyer 2003), we
presented a
theoretical analysis that shows no appreciable enhancement of the mass
outflow
rate. This conclusion is challenged by Smak (2004) who claims that equations
used in our analysis were incorrect and that in systems with short orbital
periods
substantial enhancement could occur.  In this letter, we examine the origin
of such divergent conclusions. We show that Smak's solutions are
unacceptable
from the standpoint of the equation  of continuity and that our analysis is
an
appropriate one to treat this problem.
\keywords{ accretion, accretion disks -- binaries: close --
novae, cataclysmic variables -- stars: dwarf novae}
}

\titlerunning {Enhanced mass transfer by irradiation of the secondary ?}
\maketitle

\section{Introduction}

It is well known in theory of outburst mechanisms of dwarf novae that two
rival
models (i.e., the disk instability model and the mass-transfer burst model)
have
been competing with each other. The disk instability model is now favored
because
of substantial evidence both in observations and theory and is widely
accepted
as the correct mechanism (see, e.g., monographs by Warner 1995, and
Hellier 2001). However, claims for evidence of enhanced mass overflow due to
irradiation of the secondary stars during outbursts of dwarf novae, in
particular
during the superoutbursts of SU UMa stars, still appear in the  literature
from time
to time. We have critically examined the observational evidence (Osaki and
Meyer
2003) and concluded that it is not well substantiated. Furthermore, we have
presented a theoretical analysis which shows that irradiation during
outburst
should not affect the mass transfer rate. However, this theoretical analysis
has
recently been challenged by Smak (2004) who claims that our equations
were incorrect.
The purpose of this letter is to examine critically these two papers and to
clarify
what the correct approach to treat this problem is.

\section{The problem}
During an outburst of a dwarf nova
the surface of the secondary star is irradiated by radiation
from the central white dwarf,
the inner part of the accretion disk, and the boundary layer, and the
temperature
of irradiated regions of the secondary star is raised. However, the inner
Lagrangian
point from which mass overflow occurs and the equatorial region of the
secondary
star are shielded by the accretion disk and only the region of higher
latitude is
exposed. Unequal irradiation
then causes fluid flow over the surface of the secondary. The question
is
whether this flow is capable to transport heat to the inner Lagrangian
point and thereby significantly enhance
the mass overflow rate.

Controversy has arisen between Osaki and Meyer (2003) and Smak (2004) over
how
to calculate this fluid flow. Osaki and Meyer (2003) treated it as a steady
geostrophic
flow using the Eulerian equation of fluid motion while Smak (2004) instead
calculates
trajectories of fluid particles for a given initial condition using the
Lagrangian
equation of motion. Although there are several minor differences between the
two papers such as treatment of viscosity, coordinate
systems used, and temperature of the irradiated part of the surface,  they
are not
essential and thus we concentrate here on the fundamental difference in the
treatment of the hydrodynamic flow.

\section{Fluid dynamics in a rotating frame of
reference}

The basic equations of fluid flow in a frame of reference rotating with the
binary
system (see Pedlovsky 1982) are the equation of motion,
\begin{equation}
\frac{d {\bf v}}{dt}=-\frac{\nabla P}{\rho}-\nabla \Psi-2\Omega \times {\bf
v}
+\frac{\mathcal{F}}{\rho},\label{eq:E1}
\end{equation}
and the equation of continuity,
\begin{equation}
\frac{d \rho}{dt}+\rho \nabla \cdot {\bf v}=0,\label{eq:E2}
\end{equation}
where $\bf v$ is the flow velocity in the corotating frame, $P$ and $\rho$
are
pressure and density, $\Psi$ is the total gravitational ("Roche") potential
with
contributions from both binary stars and the centrifugal force,
$\Omega$ is the angular velocity of rotation of the binary system,
$\mathcal{F}$ is the
viscous force, and symbols with bold-face signify vectors. The Lagrangian
time
derivative $\frac{d}{dt}$ follows a particular fluid element and is related
to the
Eulerian derivative $\frac{\partial}{\partial t}$ by
\begin{equation}
\frac{d}{dt}=\frac{\partial}{\partial t}+ ({\bf v}  \cdot \nabla).
\end{equation}
The equation of continuity is rewritten in Eulerian form as
\begin{equation}
\frac{\partial \rho}{\partial t}+\nabla \cdot (\rho {\bf v})=0.\label{eq:E4}
\end{equation}

In unperturbed state in which irradiation is not yet applied, we assume that
the secondary star rotates synchronously and is in hydrostatic
equilibrium so that
\begin{eqnarray}
{\bf v}_0=0 \qquad {\rm and} \qquad
\frac{\nabla P_0}{\rho_0}=-\nabla \Psi,
\end{eqnarray}
where  subscript $0$ denotes unperturbed equilibrium values.

\section{The two approaches}
We now consider the effect of irradiation. As discussed above, the
equatorial zone
of the secondary star is shadowed by the accretion disk and only a region of
higher
latitude on the secondary star is heated by irradiation. The resulting
pressure difference between these two regions drives a fluid flow. Since
the depth of the zone affected by irradiation (which we denote by $D$)
is limited to a sub-surface zone a few scale heights deep,
we have $D\ll R$, where $R$ is the stellar radius. We may then consider
flow in this surface zone only. The flow is
essentially horizontal, the ratio of the horizontal to the vertical
component of velocity is of the order of $D/R$ as seen from the continuity
equation.  The treatment of this flow is completely different in Osaki and
Meyer (2003) and in Smak (2004). We examine each of these two approaches.

\subsection{Osaki and Meyer (2003)}
In Osaki and Meyer (2003) we assume a steady geostrophic
flow over the surface of the secondary star. Since the pressure gradient
force caused by uneven heating is directed in latitudinal direction
toward the equator and since in geostrophic approximation the Coriolis
force balances the pressure gradient, the flow is zonal, i.e., directed
parallel to
the equator and, for an
inviscid fluid, the equator-ward velocity component is zero. Osaki and
Meyer (2003) also considered effects of a possible turbulent viscosity but
concluded that the equator-ward flow velocity is so slow
that no effective enhancement of the mass overflow rate is expected.

Osaki and Meyer discussed the particular case of the 2001 outburst
of WZ Sge. Basic quantities there are the 80 min orbital period of the
binary and a half thickness of $10^{9.4}$cm for the equatorial shadowed
belt. When an outburst occurs and the higher latitude region of the
secondary star is heated by irradiation, a flow sets in. Its initial
phase
may be rather complicated but it will  settle in an essentially  steady
state because the typical time-scale of an outburst is of the order of a few
days, much longer than the rotation period of the secondary star.

The order of magnitude of the left hand side of equation (1) is
given by
\begin{equation}
\frac{d {\bf v}}{dt}=\frac{\partial {\bf v}}{\partial t}+ ({\bf v}  \cdot
\nabla){\bf v}=0\left( \frac{V}{\tau}, \frac{V^2}{L}\right),
\end{equation}
where $V$, $L$, and $\tau$ are characteristic values of velocity,
length,
and time-scale of the flow. The ratio of the inertial term to
the Coriolis force on the right hand side is the so-called Rossby
number, $Ro$,
\begin{equation}
\mid \frac{d\bf v/dt}{2\Omega \times {\bf v}}\mid=0\left(
\frac{1}{2\Omega_{\perp}\tau}, \frac{V}{2\Omega_{\perp} L}\right),
\end{equation}
where $\Omega_{\perp}$ is the component of the angular velocity
perpendicular to the local surface.

In our case the Rossby number given by $Ro=1/(2\Omega_{\perp}\tau)$ is
much less than one except for the equatorial region. We may
estimate
$Ro\sim 0.01$ for the particular case considered by Osaki and Meyer in which
$\Omega_{\perp}\sim 10^{-3.15}{\rm s}^{-1}$ and $\tau\sim 10^5$s because
we are interested in the flow system on the surface on a time-scale of
days. In such a case we may neglect the inertial term on the left hand side
of the equation of motion with respect to the Coriolis term. The equation
of motion for an inviscid fluid then is
\begin{equation}
2\Omega \times {\bf v}=-\frac{\nabla P}{\rho}-\nabla \Psi.
\end{equation}

Let us use a local Cartesian coordinate system with coordinates $x$ in the
direction of constant latitude (eastward), $y$ in the direction of constant
longitude (northward), and $z$ normal to the equipotential surface
(vertically upward).  Since the main gradients are perpendicular to the
shadow band, we neglect gradients in the $x$ direction and obtain
(Osaki and Meyer 2003)
\begin {eqnarray}
-\frac{\partial P}{\partial y} -2\rho v_x\Omega_\perp&=& 0,
\nonumber \\
2\rho v_y \Omega_\perp  &=& 0.
\end {eqnarray}
Without viscosity we have the``geostrophic'' flow along the shadow band
\begin {eqnarray}
v_x &= & -\frac{1}{2\Omega_\perp} \frac{1}{\rho} \frac{\partial P}
{\partial y} \approx
-\frac{1}{2\Omega_\perp}
\frac{\Delta c^2}{l}
\nonumber \\
v_y &= & 0
\end {eqnarray}
where $c=(P/ \rho)^{1/2}$ is the isothermal sound velocity and
$l$ is the half width of the shadow band. Thus the flow velocity is at right
angle to the driving pressure gradient which is balanced by the Coriolis
force, similar to flows in weather systems and oceans on the rotating
earth. Thus without friction no heat can be advected into the shadow region
below $\rm{L_1}$. The order of magnitude of the zonal flow generated by
the geostrophic balance in equation (10) is $V\sim 10^4$cm/s
where we used $\Delta c^2\sim 10^{10.6}$ cm$^2$/s$^2$ and $L\sim 10^{9.4}$
cm, which yields a corresponding Rossby number
$Ro=V/(2\Omega_\perp L) \sim 0.003$,
confirming the self-consistency of the geostrophic flow.

Here we may add the following notes.
As in our previous paper (Osaki and Meyer 2003), we used above
a local Cartesian coordinate system to simplify the complex Roche geometry.
However, the detailed geometry is not important in the geostrophic
approximation because the geostrophic flow is directed perpendicular
to the pressure gradient and as the latter is directed from
the shadowed to the irradiated region the flow direction is along
the shadow boundary.
Flow over the surface of a mass-losing
secondary star in a semi-detached binary was already discussed by
Lubow and Shu (1975) who argued that the horizontal flow is parallel
to the isobars calling the approximation ``astrostrophic"  rather
than ``geostrophic". Our equation (10) is exactly the same as
their equation (66) in their section V.
Oka et al (2002) made numerical simulations of the surface flow
of a gas-losing secondary star and their numerical simulations
basically confirmed Lubow and Shu's astrostrophic wind.

\subsection{Smak (2004)}

Smak (2004) calculated the motion of
fluid elements
using the Lagrangian equation of motion. He begins the calculation at
the irradiated side of the shadow boundary (which we call "the starting
point") and then follows the motion of each fluid element
along its trajectory as a one-dimensional initial-value problem. The
fundamental objection to this approach is that
it does not describe a continuous hydrodynamic flow but only
calculates the motion of a free particle that abruptly starts
at a particular location following the force of a pressure gradient
obtained from cooling of the flow. Although unphysical, let us examine
Smak's approach in more detail.

\subsubsection{Quasi-steady case}
Smak (2004) apparently considers a  quasi-steady state when
in his equation (14 ) he calculates the temperature gradient from its
Lagrangian time derivative:
The complete Lagrangian derivative of temperature $T$ is
\begin{eqnarray}
\frac{dT}{dt}
& = & \frac{\partial T}{\partial t}+{\bf v}\cdot \nabla T
=  \frac{\partial T}{\partial t}+v \frac{dT}{ds}\nonumber
\\
& = &\frac{\partial T}{\partial t}+v_{x_*} \frac{\partial T}{\partial x_*}
+v_{y_*} \frac{\partial T}{\partial y_*},\label{eq:E11}
\end{eqnarray}where $s$ is the distance from the starting point
measured along the trajectory and $x_*$ and $y_*$ are local Cartesian
coordinates
used by Smak with $x_*$ directed toward the equator and $y_*$
directed in longitudinal direction. In Smak's formulation the
first and third term of the right hand side are neglected so that
\begin{equation}
\frac{dT}{dt}=v_{x_*} \frac{\partial T}{\partial x_*},
\end{equation}
i.e., steady state is assumed. Accordingly, in his Figure 2 the
trajectories of fluid elements shown are considered as stream
lines of
the flow. The third term on the right hand side in equation (\ref{eq:E11})
was neglected by Smak because he assumed $\partial T/\partial
y_*=0$ as his stream lines were fairly parallel.

We therefore adopt the assumption of steady state to further examine Smak's
solutions.  Taking the scalar product of the equation of motion
(Eq. (17) of Smak) with the velocity ${\bf v}$, we obtain
\begin{eqnarray}
{\bf v}\cdot \frac{d{\bf v}}{dt}=-\frac{ {\bf v}\cdot \nabla P}{\rho} -{\bf
v}\cdot \nabla \Psi. \label{eq:E13}
\end{eqnarray}

Using the equation of state
$P=\left(\Re\rho T/\mu\right)$,
with $\Re$ and $\mu$ gas constant and mean molecular weight,
respectively,
equation (\ref{eq:E13}) becomes
\begin{eqnarray}
{\bf v}\cdot \frac{d{\bf v}}{dt}=-\frac{\Re T}{\mu} \frac{ {\bf v}\cdot
\nabla \rho}{\rho}-\frac{\Re}{\mu} {\bf v}\cdot \nabla T-{\bf v}\cdot \nabla
\Psi.\label{eq:E14}
\end{eqnarray}

Smak then assumed that the pressure gradient is mainly produced by the
temperature gradient so that a density gradient is neglected and also takes
${\bf v}\cdot \nabla \Psi=0$ because the flow is
confined to the equipotential surface. Then we obtain
\begin{equation}
\frac{1}{2} \frac{d v^2}{dt}=-\frac{\Re}{\mu} \frac{dT}{dt}.\label{eq:E15}
\end{equation}
In Smak's formulation, the Lagrangian time derivative of temperature on the
right hand side of equation (\ref{eq:E15}) is given by radiative cooling of
the surface element (his equation (16)) and he apparently
numerically integrated his equations as an initial value problem.
However, equation (\ref{eq:E15}) can be integrated analytically to yield,
along a given stream line,
\begin{equation}
\frac{1}{2} v^2+\frac{\Re T}{\mu} ={\rm const.}\label{eq:E16}
\end{equation}
This corresponds to Bernoulli's equation.

As initial condition Smak apparently assumed that the material at the
starting
point is at rest, $v=0$ at $s=0$. Otherwise one should not have
chosen the starting point at the shadow boundary.

Equation (\ref{eq:E16}) then becomes
\begin{equation}
\frac{1}{2} v^2+\frac{\Re T}{\mu} =\frac{\Re T_0}{\mu},
\end{equation}
where $T_0$ is the temperature at the starting point. We note that this
relation holds for all the trajectories for a given
set of model parameters listed in Smak's Table 2.

At first sight the above solution might look reasonable because gas at rest
at the shadow boundary is accelerated to higher velocity by the temperature
gradient produced by cooling of the flow. However, this solution has the
fatal flaw that it does not satisfy the equation of
continuity: If one were to estimate the mass flux from the
irradiated to the shadowed side by this formula,
he would come to the unbelievable conclusion of no mass flux across
the shadow boundary because the velocity there is zero.

One can demonstrate that Smak's solution of equation (16) is
applicable only as a steady-state solution for supersonic flow.
However, as in
Smak's solution matter near the shadow boundary is subsonic,
it can not be reconciled with the equation of continuity
(Eq. \ref{eq:E4}) which in steady state is
\begin{equation}
{\bf v}\cdot\nabla \rho+\rho\nabla \cdot  {\bf v}=0.\label{eq:E18}
\end{equation}
A subsonic flow behaves more  like an incompressible fluid
and the first term on the left hand side of equation (\ref{eq:E18})
becomes negligible, ${\bf v}\cdot\nabla \rho=0$, an approximation used
by Smak
in expressing pressure gradient by temperature gradient in his equation
(13).
However, Smak's accelerating flow has a non-zero velocity divergence
in contradiction to the equation of continuity (Eq. \ref{eq:E18}).
In particular, in his solution the velocity divergence
becomes infinite at the shadow boundary
from where gas at rest is accelerated to a finite velocity:
In Smak's form, equation (\ref{eq:E15}) is written as
\begin{equation}
v_{x_*}\frac{\partial v_{x_*}}{\partial x_*}=-\frac{\Re}{\mu} \frac{dT}{dt}.
\label{eq:E19}
\end{equation}
In the shadow zone the right hand side of equation (\ref{eq:E19}) is
non-zero
so that
${\partial v_{x_*}}/{\partial x_*}$ becomes infinite when $v_{x_*}=0$.

This decisive defect of Smak's solution results from his assumption that
matter
at the shadow boundary stays at rest. But there is no steady flow on
parallel equator-ward
steamlines with zero velocity at the boundary.
If instead gas has a finite velocity across the shadow boundary,
matter from higher latitude must move in to replace matter that moved into
the shadow region.
Since this matter has already a significant zonal velocity when passing
the shadow boundary,  a picture very different from that of Smak emerges.

One may ask whether at the
shadow boundary cool material might well up from
below to fill the gap created by gas drained equatorwards,
and always replace it by matter with zero zonal velocity.
However, in order for up-welling  gas to prevent inflow from higher latitude
it must provide the same pressure at the same depth as
the neighboring irradiated region.
With vertical hydrostatic equilibrium this means the
same temperature down to the same depth. To heat up a column by irradiation
to that depth requires the same time as that for which the existing
hot material was exposed i.e. the time since the start of irradiation.
As soon as this time exceeds the time on which heated
matter flows in, of the order of the sound travel time (about an hour),
up-welling of gas is suppressed by inflow.

\subsubsection{Unsteady case}

This shows that Smak's approach is unacceptable as a description of
steady flow. Let us examine whether his solutions might reflect an
unsteady initial flow. In this case we suddenly switch on the pressure
gradient and observe what happens. As Smak argued, the matter at the heated
side of the shadow boundary first moves along the pressure gradient, i.e.
equator-wards. This lowers the density at the shadow boundary and creates
a pressure deficit with respect to neighboring higher latitude regions
from which gas then moves in to fill the void. Since this gas has lower
specific angular momentum than that which already left the starting
point, it has there a significant zonal velocity. This velocity
increases with time
as material from higher and higher latitude is drained into the shadow
region.
This adjustment occurs over all the surface of the secondary star until
the flow at the boundary of the shadow region has a zonal component
sufficient for the Coriolis force to balance the pressure
gradient, and our steady geostrophic state is reached.

One may estimate the time needed to establish geostrophic equilibrium.
In the case of WZ Sge discussed above and for the minimal model
of a spherical secondary star already gas from a distance of
$10^{9.6}$cm from the equator, on reaching the interior of the shadow belt
would have enough zonal velocity to balance pressure gradient by
Coriolis force. The hydrodynamic time for this matter to arrive in
the shadow belt is again only of the order of an hour, the same order of
magnitude
as the orbital period.

\section{Conclusion}

We conclude that in dwarf novae outbursts enhanced mass transfer by
irradiation of the secondary star is prevented by the strong Coriolis force.

\end{document}